\documentclass{article}

\usepackage{arxiv}

\usepackage[utf8]{inputenc} % allow utf-8 input
\usepackage[T1]{fontenc}    % use 8-bit T1 fonts
\usepackage{hyperref}       % hyperlinks
\usepackage{url}            % simple URL typesetting
\usepackage{booktabs}       % professional-quality tables
\usepackage{amsfonts}       % blackboard math symbols
\usepackage{nicefrac}       % compact symbols for 1/2, etc.
\usepackage{microtype}      % microtypography
\usepackage{graphicx}
\usepackage{epstopdf, epsfig}
\usepackage{amsmath}
\usepackage{graphicx}
\usepackage{graphics}
\usepackage{subfig}
\usepackage{listings}

\title{Development of open source instruments for in-situ measurements of waves in ice}
\author{
  J. Rabault$^1$, J. Voermans$^2$, G. Sutherland$^3$, A. Jensen$^1$, A. Babanin$^2$, K. Filchuk$^4$ \\
  $^1$: University of Oslo, Oslo, Norway \\
  $^2$: University of Melbourne, Melbourne, Australia \\
  $^3$: Environment and Climate Change Canada, Dorval, Canada \\
  $^4$: Arctic and Antarctic Research Institute, St. Petersburg, Russia\\
}

\begin{document}
\maketitle

\begin{abstract}
The interaction between surface waves and sea ice involves many complex
    physical phenomena such as viscous damping, wave diffraction, and nonlinear
    effects. The combination of these phenomena, together with considerable
    variability in ice configuration, ranging from viscous grease ice slicks to
    large icebergs through closed drift ice and landfast ice, makes it
    challenging to develop robust and accurate waves in ice models. In this
    context, a reason for the challenges modellers are facing may lie in the
    mismatch between the relative scarcity of waves in ice data available for
    testing theories, and the wide diversity of phenomena happening at sea.
    This lack of experimental data may be explained, at least in part, by the
    high cost of waves in ice instruments. Therefore, development of open
    source, low-cost, high-performance instrumentation may be a critical factor
    in helping advance this field of research. Here, we present recent
    developments of a new generation of open source waves in ice instruments
    featuring a high accuracy Inertial Motion Unit as well as GPS, on-board
    processing power, solar panel, and Iridium communications. Those
    instruments are now being used by several groups, and their simple and
    modular design allows them to be customized for specific needs quickly and
    at reduced cost. Therefore, they may be an important factor in allowing
    more data to be gathered in a cost-effective way, providing much-needed
    data to the waves in ice community. This approach is here validated by
    presenting recent sea ice drift and wave activity data, and comparing these
    results with those obtained with commercially available buoys. In addition,
    our design may be used as a general platform for cost-effective development
    of other in-situ instruments with similar requirements of low-power data
    logging, on-board computational power, and satellite communications.
\end{abstract}

\section{Introduction}

A number of complex phenomena are involved in the mechanics of wave-ice interaction.
These include, to name but a few, viscous damping
\citep{WeberArticle, rabault_sutherland_gundersen_jensen_2017}, wave
diffraction \citep{SquireOOWASI}, and nonlinear effects in the ice
\citep{AnalysisPolarsten}. The complexity of these phenomena, their intrication
and combination in all experimental datasets, and the (relative) similarity
of their effect on waves propagating in ice (i.e., wave damping and wave-induced drift)
are key factors that explain the challenges faced by the waves in ice community
\citep{rabault2018investigation, squire2018fresh, babanin2019waves}. This inherent complexity is
unfortunate, as getting detailed understanding of wave-ice interaction is necessary
for allowing safe human activities in the polar regions, as well as improving both
weather and climate models. As a consequence of both the
challenges presented by wave-ice interaction, and the practical interests at stake,
several groups are actively working on the topic, and the field of wave-ice interaction
presents a recent surge in activity \citep{JGRC21649, Wang201090,
JGRC:JGRC11467, Zhao201571, SUTHERLAND2019111, rabault_sutherland_jensen_christensen_marchenko_2019, MARCHENKO2019101861, voermans2019wave,
doi:10.1146/annurev-fluid-010719-060301}.

A natural direction for improving both understanding of wave-ice interaction,
and our modeling abilities, is to generate more data from field measurements under a variety of waves,
ice, and currents conditions. There, several challenges have been encountered historically.
First, the polar regions are famous for their harsh climate and strong storm events,
therefore, putting demanding requirements on instruments to be deployed there. Second,
instruments able to perform waves in ice measurements have traditionally been expensive
black boxes, which both limits modularity and ease of use, and puts severe constraints
on the number of instruments that can be deployed given constrained budgets. However,
recent developments in free and open source electronics and software are opening
new opportunities for bypassing the expensive, inefficient process of acquiring
scientific instruments for waves in ice measurements from private companies. 
Such approach has notably been pioneered in a number of recent works \citep{rabault_sutherland_gundersen_jensen_2017,
RABAULT2020102955}, which introduced hardware and software that through iterative evolution have reached the point
where a fully operational wave in ice instrument is now available as Free Open Source Software
and Hardware (FOSSH).

In this paper, we offer a short overview of the FOSSH waves in ice instruments,
and we present a new dataset of both waves in ice and ice drift obtained
recently in the Antarctic. During this deployment, commercial buoys were deployed
at similar locations to the FOSSH instruments. Comparing the data from the FOSSH
versus commercial instruments, we fully validate the wave measurements and in-situ
data processing. Finally, we discuss in details the next steps to be taken for further
improving the FOSSH instruments, therefore, laying the foundations for the next
iteration of the design.

\section{Full Open Source Software and Hardware instruments}

The FOSSH instruments used in the present paper were developed over time by J. Rabault, G. Sutherland
and O. Gundersen at the University of Oslo over the period 2016-2018, and used first for a real-world,
full-scale deployment in September 2018 \citep{7513396, rabault_sutherland_gundersen_jensen_2017, RABAULT2020102955}.
Since then, one more series has been built and deployed in the context of a collaboration between the University of Melbourne, AARI and the University of Oslo, following
the design presented in \citet{RABAULT2020102955}. Two more series are awaiting
deployment. The full design is
released as FOSSH material, see: \url{https://github.com/jerabaul29/LoggerWavesInIce_InSituWithIridium}.

\begin{figure}[h]
  \begin{center}
    \includegraphics[width=.65\textwidth]{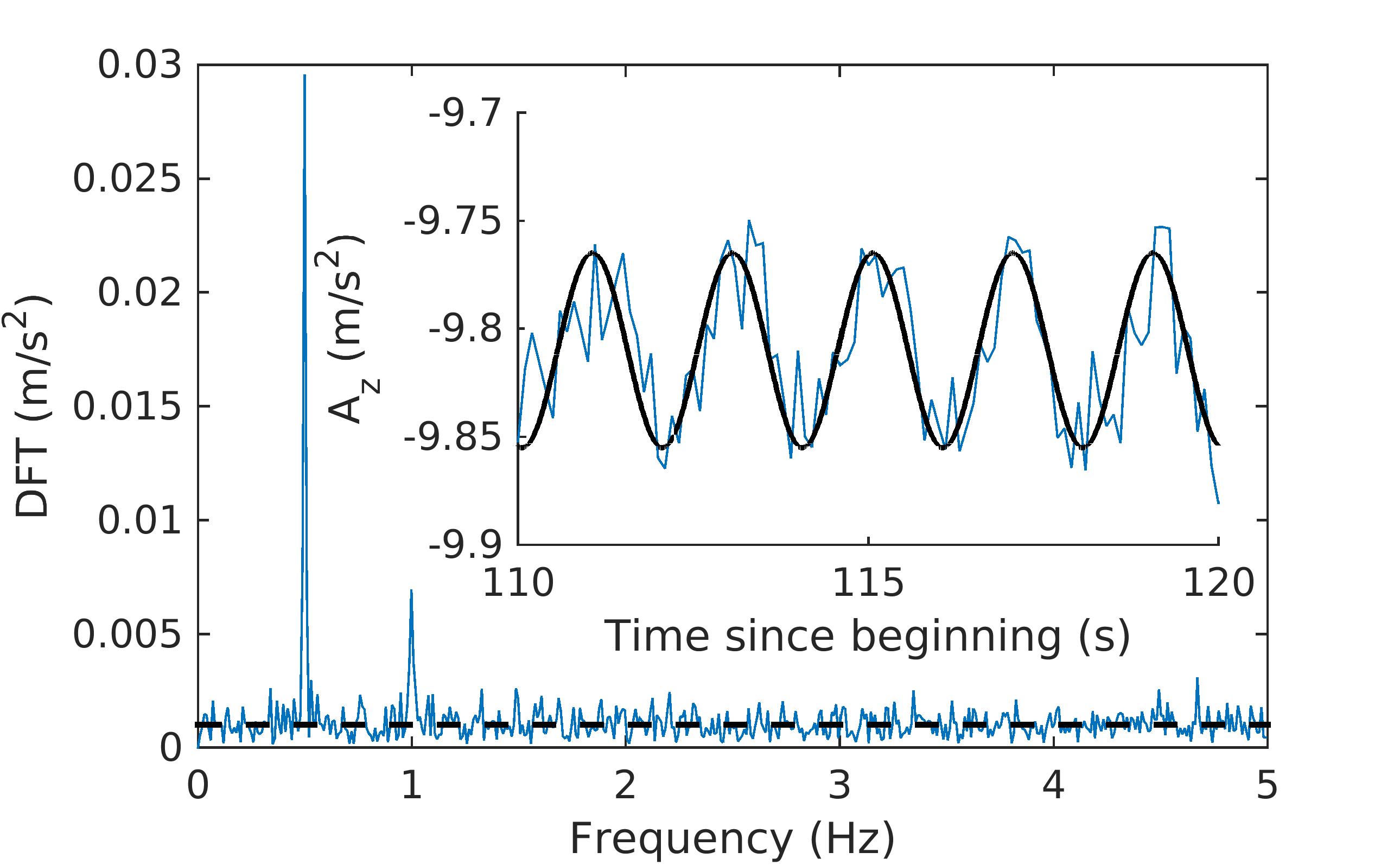}
    \caption{\label{test_lab} Test in the laboratory of the accuracy of the acceleration measurements
    provided by the VN100 Inertial Motion Unit. The VN100 IMU is positioned on a styrofoam floe in a
    wave flume, and waves of amplitude 3 mm (2 s period) are successfully recorded. Reproduced from
    \citet{7513396}. Since this work, more optimal configuration of the VN100 IMU has allowed
    to reduce even further the level of noise.}
  \end{center}
\end{figure}

The central reason allowing the development of low-cost, easy-to-build, accurate instruments lies
in the recent development and democratization of micro electronics that has been allowed by several
Open Source projects. This means that one can now, with minimal previous electronics or programming
knowledge, design, program, and assemble devices containing micro controllers, micro computers, GPS
modules, Iridium satellite communications, and powerful sensors such as the high-accuracy VN100
Inertial Motion Unit (IMU) \citep{VN100specs}. In particular, laboratory testing \citep{7513396} and
previous field deployments \citep{JGRC21649} have confirmed the ability of the VN100 to measure
waves of amplitude down to the millimeters range (see Fig. \ref{test_lab}).

\begin{figure}[h]
  \begin{center}
    \includegraphics[width=.45\textwidth]{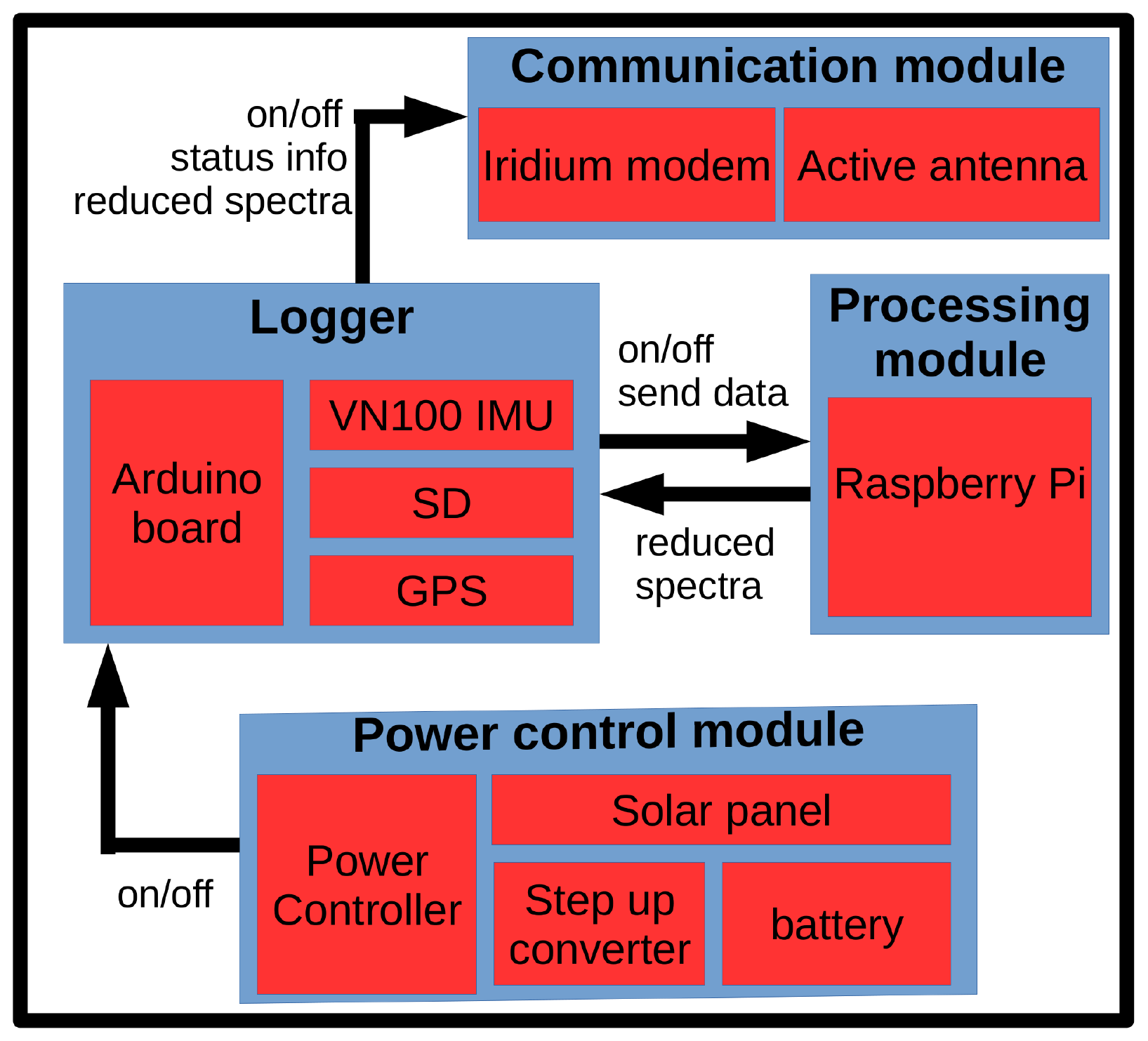}
    \caption{\label{fig_modules} The instruments are built following a modular design. The different modules can be combined in a flexible way to implement a wide variety of in-situ data acquisition, data processing, and satellite communications. Reproduced from \citet{RABAULT2020102955}.}
  \end{center}
\end{figure}

While more details are available in a recent paper \citep{RABAULT2020102955}, we
summarize here the main lines of the FOSSH design for the sake of completeness. The instrument
features 4 main hardware components: a power control module, a micro-controller based
logger, an on-board microcomputer, and a satellite communication unit (see Fig. \ref{fig_modules}). Each of these
modules is built using inexpensive, off-the-shelf components. The combination
of these modules constitutes a general purpose, flexible, cost-efficient platform
from which instrument designs can be easily and quickly iterated.

\begin{figure}[h]
  \begin{center}
    \includegraphics[width=.45\textwidth]{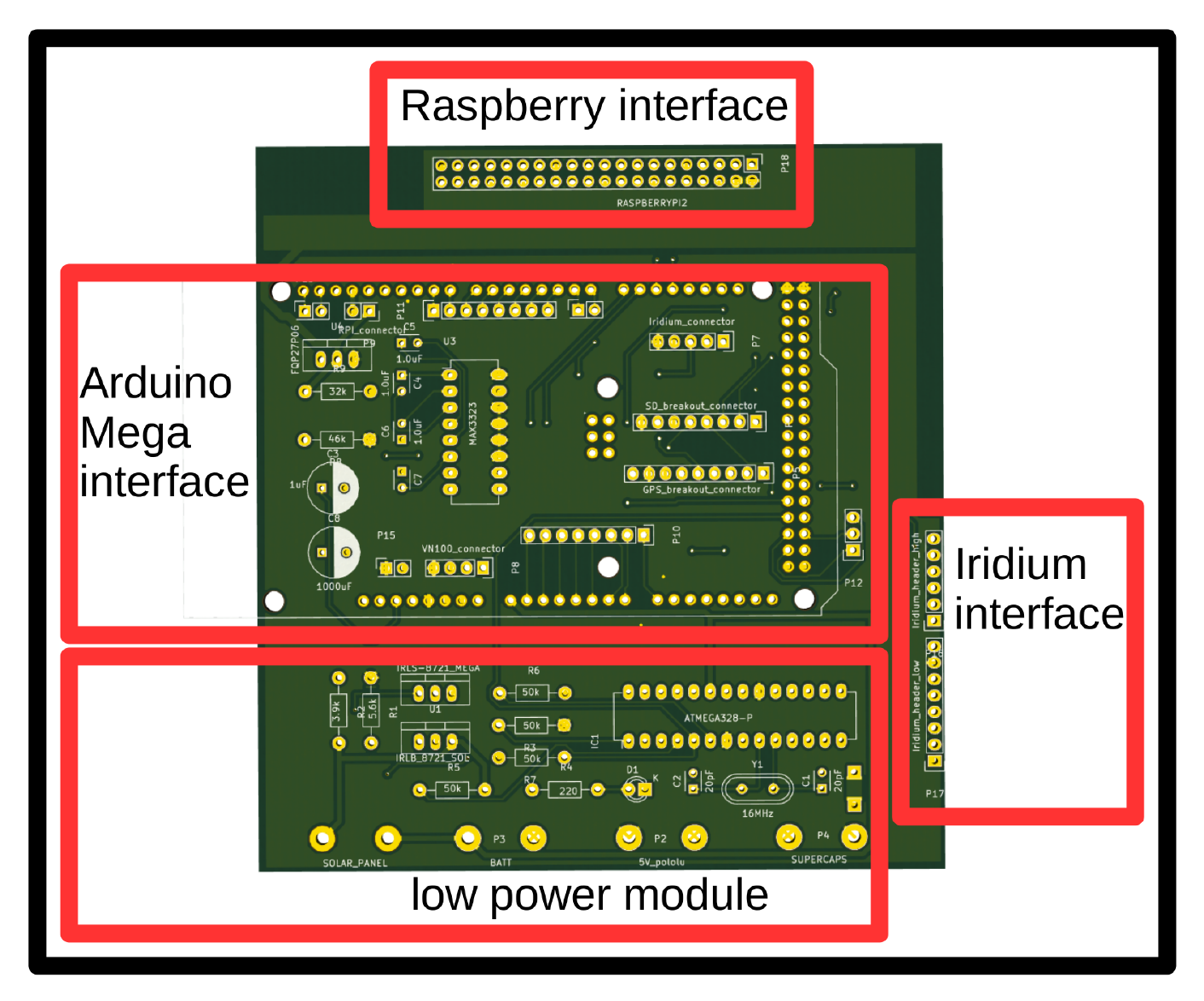}
    \caption{\label{fig_PCB} The instruments are organized around a single main PCB design, which is part of the fully released FOSSH material. Reproduced from \citet{RABAULT2020102955}.}
  \end{center}
\end{figure}

More practically, the instruments are built based on a single Printed Circuit Board (PCB),
available together with the software, on which common components are soldered (see Fig. \ref{fig_PCB}).
The PCBs can be inexpensively ordered from any producer, and the design choices were
made so as to maximize the ease-of-assembly, so that any person with minimal soldering skill
can mount all the components needed in typically no more than about 3 hours per instrument
built.

To summarize, the FOSSH instruments have considerable advantages compared to commercially produced instrumentation:

\begin{itemize}
  \item \textbf{Price}: the cost of the FOSSH used in this study is around 2K USD. This is around 3-50 times less than commercial alternatives,
  which are commonly sold for 5K-100K USD. Though it should be noted that the FOSSH requires a few hours time for assembly
  and testing, and additional time to familiarize with the hardware and software specifications, the FOSSH is designed
  from commonly used software and off-the-shelf hardware thereby limiting the specific expertise required (particularly
  for academics) to build and work with the instrument
  
  \item \textbf{Flexibility}: as the complete instrument design is open-source, all aspects of the instrument can be altered to
  facilitate owner and project needs. Commercially-built instruments will always have a certain degree of hidden software
  and perhaps hardware that limits the full understanding of the instrument functioning. Both the software and hardware of
  the FOSSH are fully customizable, allowing changes in data recording, transmission and processing algorithms. This also allows
  accommodation of additional measurements, including temperature, wind speed, atmospheric pressure, by performing minor updates
        to the hardware and software.
  
  \item \textbf{Technology type}: the FOSSH is currently based around a leading high-performance Inertial Motion Unit (IMU). Though this
  is the most expensive component of the FOSSH by far, it allows for wave motion accuracies down to a few millimeters. 
\end{itemize}

\begin{figure}[h]
  \begin{center}
    \includegraphics[width=.45\textwidth]{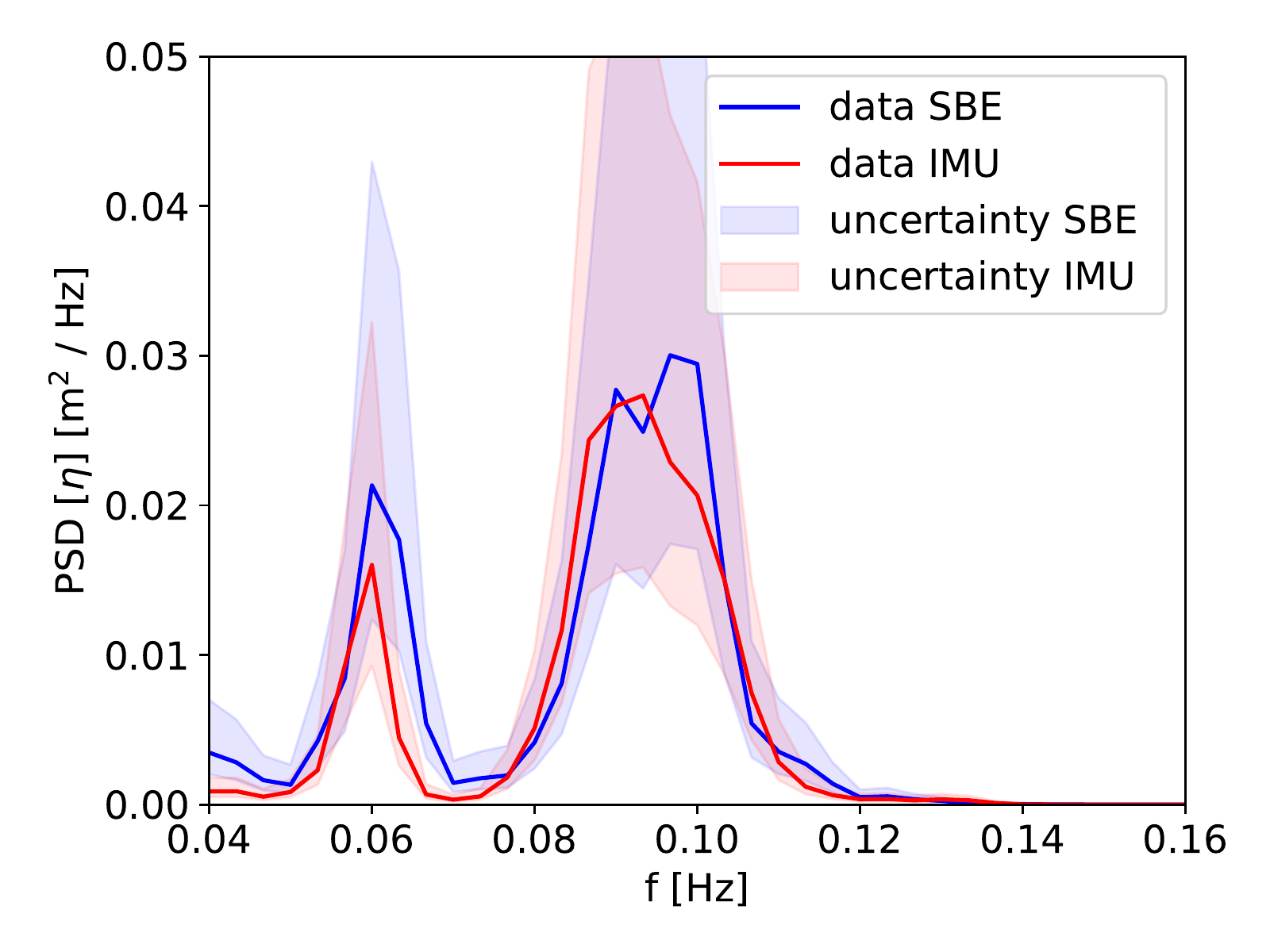}
    \caption{\label{validation_pressure} Validation of the open source instruments (relying on a VN100 IMU) against
    pressure sensor measurements of waves (SBE spectra). Reproduced from \citet{RABAULT2020102955}.}
  \end{center}
\end{figure}

The first real-world deployment of the instruments took place in 2018 \citep[described by ][]{RABAULT2020102955},
and allowed to both fully test the instruments at sea, validate the results against a measurement of waves
using a pressure sensor (see Fig. \ref{validation_pressure}), and collect a first dataset that is
currently under further analysis.

Following this first deployment, new series of instruments were built as previously named, and
we will here present recent data from a measurement campaign during which two Sofar Spotter wave buoys were deployed close to FOSSH instruments
in the Antarctic. The Spotter
buoys are commercial wave buoys (\url{https://www.sofarocean.com/products/spotter}), and have been
extensively tested against the industry-standard Waverider buoys \citep{doi:10.1175/JTECH-D-18-0151.1} and NDBC weather stations \citep{voermans2019estimating}. Though the Spotter wave buoys were
initially deployed with the idea to measure ocean waves after the sea ice on which they were deployed would have disappeared, they provide a valuable opportunity here to further validate
the performance of the FOSSH instruments.

\section{Recent deployment in the Antarctic, and validation against the Spotter buoy}

\begin{figure}[h]
  \begin{center}
    \includegraphics[width=.65\textwidth]{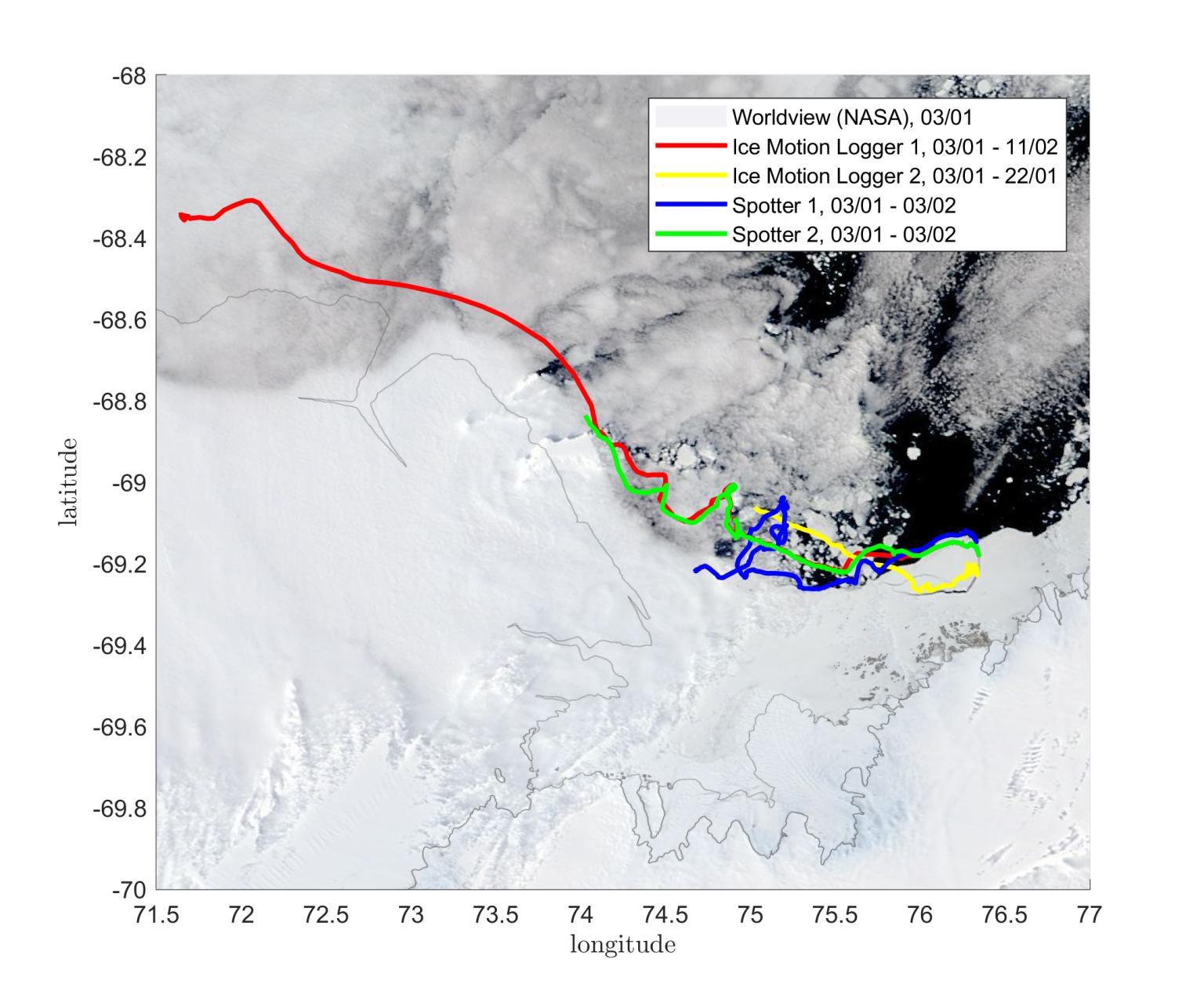}
    \caption{\label{trajectories} Illustration of the trajectories of the different instruments (satellite imagery on 03-01, Worldview NASA).
    Deployment took place on what was initially landfast ice. Following incoming waves, the ice broke
    up after around 1 month of deployment, and the instruments started to drift. As of the middle of
    February 2020, i.e. after around 2 months of deployment, only a single FOSSH instrument was still
    active. It is believed that the other instruments were crushed due to the combination of ice
    breakup and incoming waves.
    }
  \end{center}
\end{figure}

Data from the Antarctic sea ice cover were obtained during the second measurement campaign using FOSSH buoys (the first campaign
took place in September 2018 in the Arctic and was reported in \citet{RABAULT2020102955}). The deployment
took place in December 2019 on landfast ice in Antarctica (approximate deployment location: 69.25S, 76E).
This deployment included two FOSSH instruments, as well as two Spotter wave buoys. While initially the ice remained
still and attached to land, around one month after deployment the ice started breaking (see Fig. \ref{trajectories}). The ice breakup
happened close to the location of the instruments (both FOSSH and Spotter buoys). As a result, the instruments
got separated and the trajectories
of the different instruments diverged over time. After a few weeks, one FOSSH instrument (22-01) and then the two Spotter buoys (03-02)
stopped transmitting, probably due to the combination of ice breakup and incoming waves crushing the instruments housings.

\begin{figure}[h]
  \begin{center}
    \includegraphics[width=.65\textwidth]{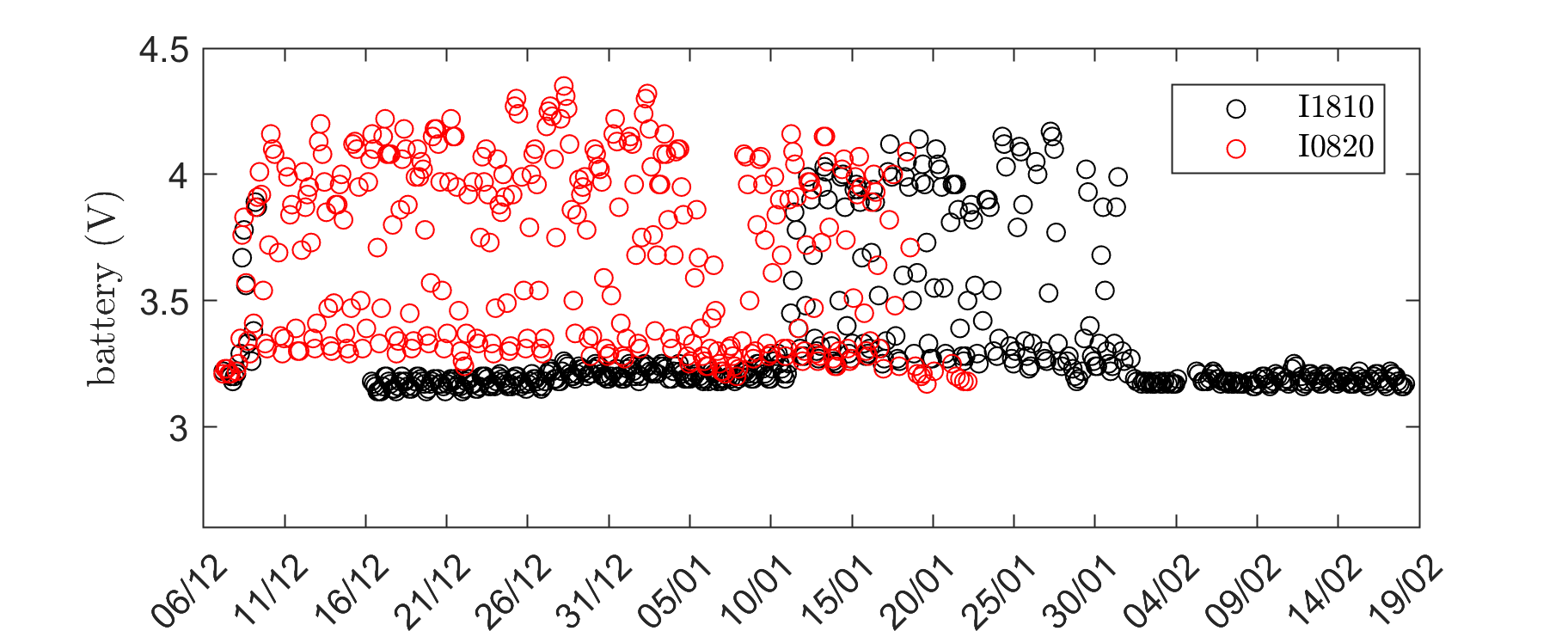}
    \caption{\label{battery_levels} Time series of the battery levels reported by the FOSSH instruments. Thanks to the
    solar panel and low power design, the batteries stayed fully charged during the whole period of the
    deployment. What is believed to be a slight timing issue in the code or a bad contact during battery assembly
    inside the logger has caused intermittent battery readout issues on the instruments, which should be fixed
    in future versions. However, despite this issue, it is clearly visible that the batteries remain more or less
    fully charged (around 3.3V) during the whole deployment, thanks to efficient power management and the presence
    of a small 3.5W solar panel on top of the pelicase housing the electronics.
    }
  \end{center}
\end{figure}

To confirm the hypothesis that the termination of transmission of at least one of the FOSSH instruments was due
to crushing by the ice, battery levels of the FOSSH devices can provide guidance. Similar analysis over a
deployment that lasted for about 2.5 weeks until loosing contact indicated that no significant battery level
drop could be observed \citep{RABAULT2020102955}. Here, we observe the same phenomenon, i.e. the batteries remained
fully charged due to favorable weather conditions providing near-continuous solar panel input (Fig. \ref{battery_levels}). It should be noted that an
unknown minor technical issue created some spread in the battery voltage reported. We believe that this is a benign
artifact probably due to some low level issue on the firmware side, and we are working towards resolving it. Nevertheless, the long deployment duration confirms
that the solar panel is sufficiently powerful to allow for unlimited operation when solar input is
provided.

\begin{figure}[h]
  \begin{center}
    \includegraphics[width=.65\textwidth]{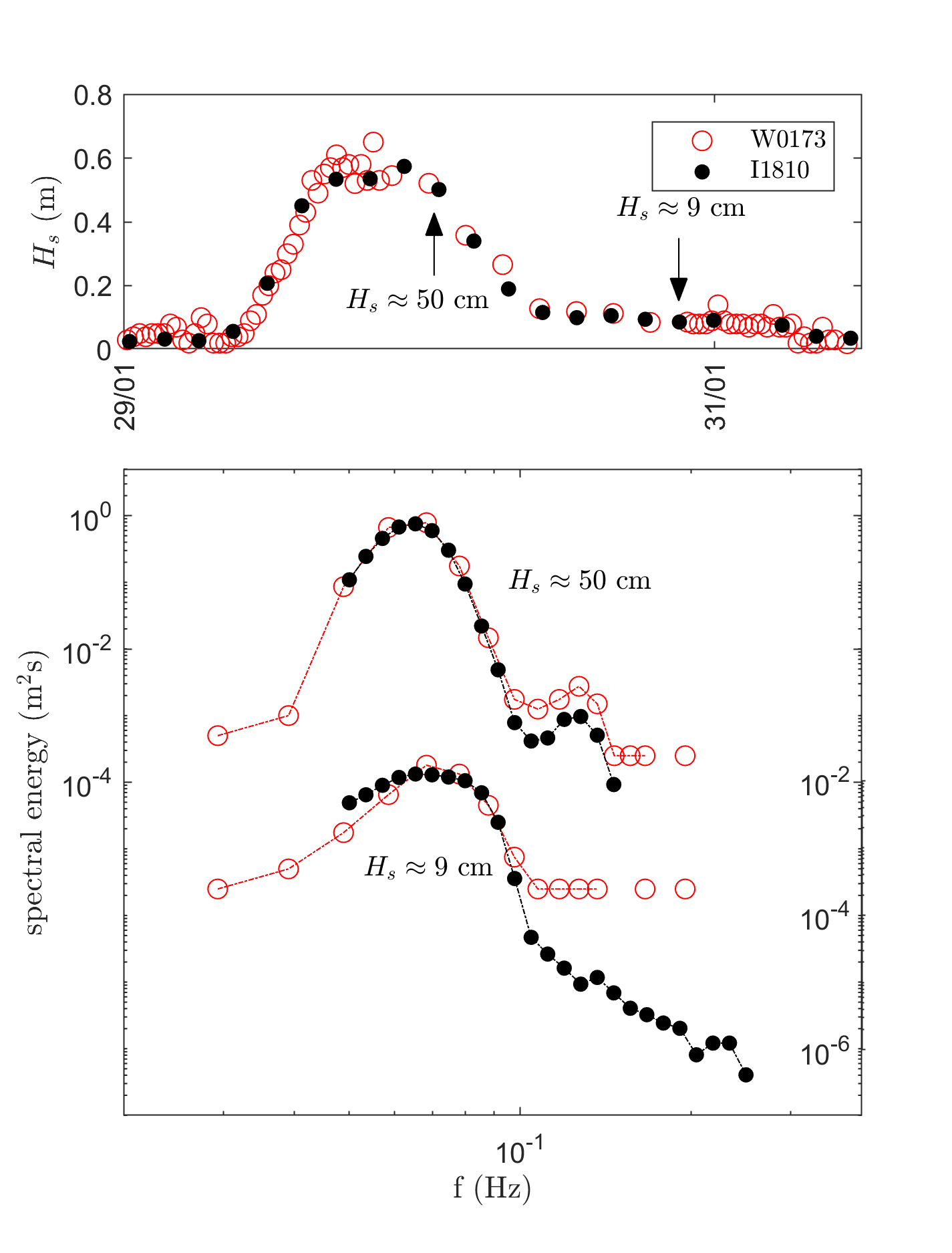}
    \caption{\label{wave_results} Comparison between the wave measurements obtained from the
    Spotter wave buoy (W0173) and the FOSSH waves in ice instrument (I810) that were closest to
    each other (see trajectories in Fig. \ref{trajectories}). Top: comparison of the Significant
    Wave Height (Hs) reported by the two instruments over the period of time 29-31 January 2020.
    Both instruments report very similar results. Bottom: wave spectra reported at two different
    times (corresponding to Hs = 50 cm and Hs=9 cm, respectively, as highlighted on the top panel).
    The location and height of the peaks are very similar for both instruments. The second pair of
      spectra (Hs=9cm) has been slightly shifted vertically for ease of representation, and should
      be interpreted with the Y-label on the right edge of the figure.}
  \end{center}
\end{figure}

As visible in Fig. \ref{trajectories}, the trajectories of one of the FOSSH instruments
and one of the spotter buoys remained close for an extended amount of time. Over this period
of time, several significant wave events took place, which were recorded by both instruments.
A summary of these measurements is provided in Fig. \ref{wave_results}. As visible there,
the time series for the significant wave height indicate excellent agreement between both
instruments. Similar quantitative agreement is observed on the frequency spectra, as is
also visible on Fig. \ref{wave_results}. While minor differences are observed at the high-frequency end of the spectra,
the energy in this range is considerably below the wave energy levels that can be reasonably resolved at these
frequencies by these instruments. In addition, this difference may be explained by both artifacts
in the processing used (for instance, the FOSSH instruments rely on a well-tested taper function for windowing
when taking Fourier transform, and uses the Welch method), and the different sensors used to measure the motion.

\section{Conclusion}

In recent years, Fully Open Source Software and Hardware (FOSSH) instruments are
becoming a promising avenue for gathering critical data in challenging remote environments. This,
in turn, holds many promises for the waves in ice community. In
this proceedings, we presented a summary of the technical results obtained so far with one
such family of FOSSH instruments. They are shown to perform reliably in demanding
polar conditions, and to compare satisfactorily with commercially available devices.
Therefore, the barrier to entry for new actors who would want to participate in
monitoring of ice drift, breakup, as well as wave propagation in the ice, is drastically reduced,
and established groups can get access to new low-cost alternatives to traditional
instrumentation solutions.

While several successes were obtained, this is by no means an end to the gradual
improvements of our FOSSH instruments design. As new electronics are regularly released,
we expect to be able to further improve the design so as to continue simplifying the
assembly of the instruments, reduce their cost, and improve even further both power efficiency and
accuracy of the measurements. We hope that these efforts may inspire new groups to
join this collaborative project and contribute with their technical abilities. In the
near future, we will be working towards incorporating additional measurements
such as temperature, pressure, and hygrometry, in a standardised way, and we will as well
investigate the possibility to monitor ice breakup by including acoustic sensors.

\bibliographystyle{jfm}
\bibliography{template}

\end{document}